\documentclass[preprint,12pt,authoryear]{elsarticle}

\usepackage{amssymb}

\journal{Microelectronics Reliability}

\usepackage[latin1,utf8]{inputenc}   
\usepackage[T1]{fontenc}        
\usepackage[ngerman,english]{babel}     
\usepackage{amsmath,amssymb}    
\usepackage{upgreek} 
\usepackage{amsthm}
\usepackage{graphicx}
\usepackage{listings} 
\usepackage{algorithm2e}
\usepackage[makeroom]{cancel}
\usepackage{longtable}
\usepackage{dsfont} 
\usepackage{graphicx}
\usepackage[svgnames]{xcolor}
\usepackage{lipsum}
\usepackage{tabularx}
\usepackage{epigraph}
\usepackage[toc,page]{appendix}
\usepackage{hyperref} 
\usepackage{mathtools} 
\usepackage[symbol]{footmisc}

\hypersetup{ 
	colorlinks,
	citecolor=black,
	filecolor=black,
	linkcolor=black,
	urlcolor=black
}
\allowdisplaybreaks 

\usepackage{setspace} 
\theoremstyle{definition}
\newtheorem{definition}{Definition}

\theoremstyle{remark}

\begin{document}

\begin{frontmatter}

\maketitle



\author[add1]{Lukas Sommeregger\corref{cor1}}
\ead{lukas.sommeregger@infineon.com}
\author[add2]{J\"urgen Pilz}
\ead{juergen.pilz@aau.at}
\cortext[cor1]{Corresponding author}

 \address[add1]{Infineon Technologies Austria AG}
 \address[add2]{Universit\"at Klagenfurt, Department of Statistics}
 
\title{Quality Control of Lifetime Drift in Discrete Electrical Parameters in Semiconductor Devices via Transition Modeling}

 


\begin{abstract}

Semiconductors are widely used in various applications and critical infrastructures. These devices have specified lifetimes and quality targets that manufacturers must achieve. Lifetime estimation is conducted through accelerated stress tests. Electrical parameters are measured at multiple times during a stress test procedure. The change in these Electrical parameters is called lifetime drift. Data from these tests can be used to develop a statistical model predicting the lifetime behavior of the electrical parameters in real devices. These models can provide early warnings in production processes, identify critical parameter drift, and detect outliers.\\

While models for continuous electrical parameters exists, there may be bias when estimating the lifetime of discrete parameters. To address this, we propose a semi-parametric model for degradation trajectories based on longitudinal stress test data. This model optimizes guard bands, or quality-guaranteeing tighter limits, for discrete electrical parameters at production testing. It is scalable, data-driven, and explainable, offering improvements over existing methods for continuous underlying data, such as faster calculations, arbitrary non-parametric conditional distribution modeling, and a natural extension of optimization algorithms to the discrete case using Markov transition matrices.

\end{abstract}

\begin{highlights}
\item Modelling semiconductor parameter drift
\item 	Statistical model for lifetime behavior
\item 	Semi-parametric model for discrete parameters
\item 	Optimizes guard bands for lifetime quality
\item 	Scalable, data-driven, and explainable model
\end{highlights}


\end{frontmatter}



\section{Introduction}\label{sec1}
The significance of semiconductors has become increasingly evident in recent years, particularly in the automotive and aviation sectors, where the reliability of electronic components is paramount to ensuring passenger and operator safety. While standard practice involves testing devices during production to guarantee quality at the time of shipping, this assessment only provides a snapshot of the device's state. To ensure quality over the device's lifetime, reliability tests are conducted, subjecting devices to stress tests and monitoring their performance over time, as outlined in, e.g., the \cite{IEC} standard.\\

Accelerated stress tests are a crucial aspect of assessing lifetime behavior, as specified in standards such as \cite{AECQ100} and \cite{AECQ101}. These tests involve measuring electrical parameters at an initial stage, subjecting devices to harsher-than-usual conditions, and measuring parameters at predetermined intervals. The resulting change in electrical parameters over time, known as lifetime drift, provides valuable insights into degradation processes within the devices. This information can be leveraged to identify root causes in productive processes and establish guard bands to guarantee lifetime quality of parts.

We propose a data-driven, non-parametric model to quantify uncertainty and lifetime drift in stress test data, focusing on discrete electrical parameters. This scalable model is applicable to various panel data applications, involving multiple time series from the same underlying process.

In semiconductor manufacturing, specification limits define the acceptable range for parameters, while test limits, chosen to meet quality criteria, account for lifetime drift. The goal is to maximize the number of shipped parts while maintaining the stringent $1$ ppm quality target, which allows only one in a million shipped devices to exceed those limits. This process, known as guard banding, involves setting test limits based on stress test data and establishing guard bands between specified and test limits \cite{Healy}, \cite{Jeong}.

Our research addresses a critical gap in modeling discrete parameters, building on previous efforts in lifetime drift modeling for continuous parameters. This work contributes to improving quality control and stability in semiconductor production processes, ultimately enhancing the reliability of electronic components in safety-critical applications.

\subsection{State-of-the-Art}

Stress tests are a common practice, often used to simulate real-life lifetime behavior, as reflected in several industrial standards, such as \cite{AECQ100}, \cite{IEC}, and \cite{AECQ101}, and discussed in \cite{st1}, \cite{st2}, \cite{st3} and \cite{Meeker}. Guard banding, a risk control process, has been extensively studied, for example in \cite{Healy}, \cite{Jeong}, \cite{Chou}, \cite{Mottonen}, \cite{gb1}, and \cite{g1}. However, the use of drift data from lifetime stress tests for guard banding has received less attention, with exceptions including \cite{Hofer} and \cite{Hofer2}, which use copula-based models, and \cite{SommereggerPaper}, which employs mixed models. An approach using machine learning (ML) methods for regularization has been proposed in \cite{SommereggerDataDriven}.

These papers represent the current state-of-the-art in guard banding lifetime drift, focusing on continuous parameters. The handling of discrete parameters remains an area of investigation. Our paper offers a new approach to address this gap.

Given the requirement for the model to run on edge devices, it must be computationally and storage-efficient. The data structure from stress testing is highly time-censored, emphasizing the need for accurate guard banding and tail density estimation of exceedance probabilities. This renders classical panel data analysis methods, such as those in \cite{Singer}, \cite{RegressionF}, or \cite{BayesianCluster}, unsuitable for the task at hand.

For continuous parameters, multivariate longitudinal data modeling approaches exist, such as outlined in \cite{Liang}, \cite{lgpr}, and \cite{TLongMiss}. However, the extension to discrete parameters remains inconclusive. Non-parametric and transition-based approaches typically focus on continuous data, like \cite{MLNP}, and Sklar's theorem does not hold for discrete data \cite{Geenens}, limiting the applicability of classical copula methods.

\begin{figure}\centering
	\includegraphics[width=.6\linewidth]{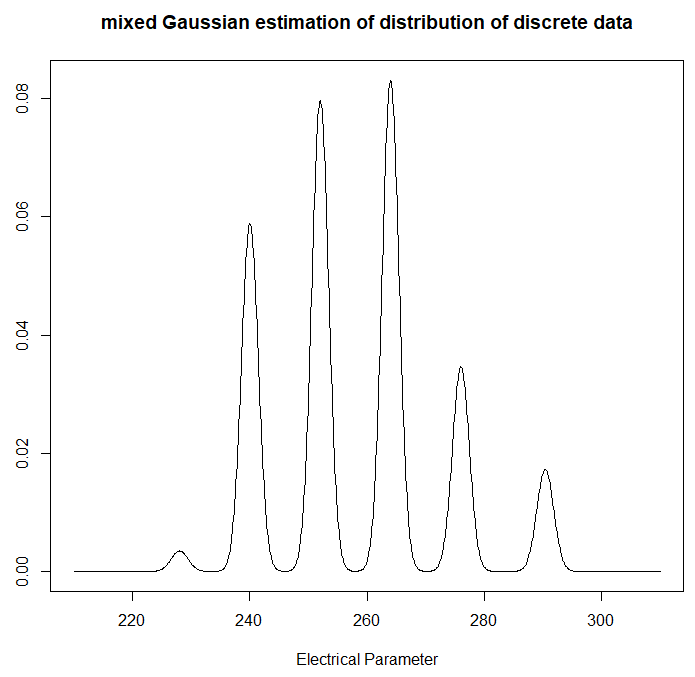}
	\caption{The Gaussian mixed estimation may lead to overfitting of the distribution on discrete data.}
	\label{gaussmix}
\end{figure}

\begin{figure}\centering
	\includegraphics[width=.6\linewidth]{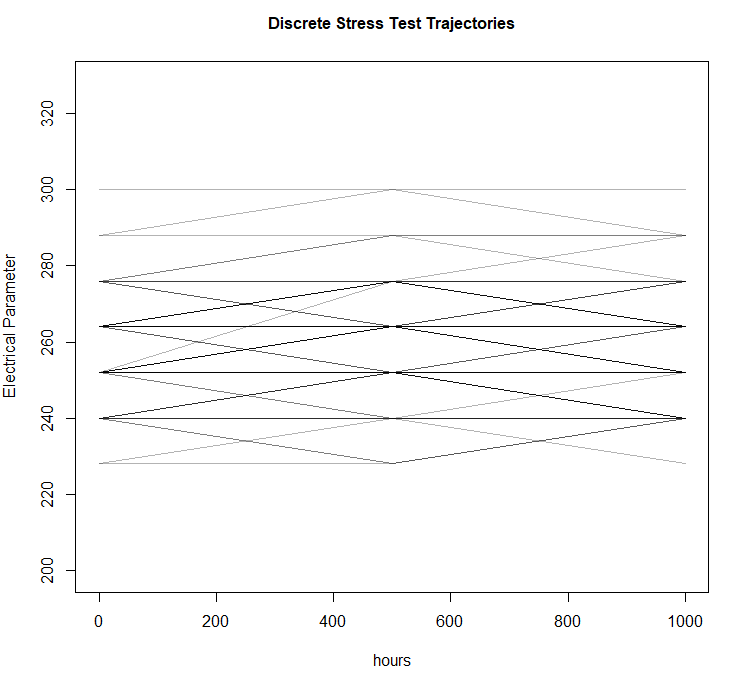}
	\caption{An example of discrete trajectory data. Continuous mixed modeling may misrepresent the discrete distribution.}
	\label{data_01}
\end{figure}

No proposals have been made for discrete counterparts to the continuous multivariate distribution-based models, as seen in \cite{Hofer} and \cite{SommereggerPaper}, to the best of the authors' knowledge. Continuous models may encounter difficulties when dealing with certain types of discrete parameter data, such as when using mixed multivariate distribution modeling, which may lead to a misrepresentation of the underlying distribution, as demonstrated in Figure \ref{gaussmix}, where a mixed model was fitted to the longitudinal data from Figure \ref{data_01} and the distribution of the parameter at the last time-step was plotted .


To address this research gap, we propose a semi-parametric Markov-chain-based transition model for lifetime drift that deals with the sparse data obtained from stress testing and correction for tester offset. We further show how to use the model to naturally obtain interpolated distributions of lifetime drift behavior between readout times.\\
The proposed model fills a space of flexible and easy-to-compute distributional modeling of multivariate discrete data. One main benefit of the method is that it provides a copula-free way of modeling skewed conditional distributions and transition matrices which allow for easy optimization and computation of guard bands. A preview of this work was given at \cite{ICSTA}.

\section{Data Structure and Preparation}\label{Struct}

\begin{figure}\centering
	\includegraphics[width=1.1\linewidth]{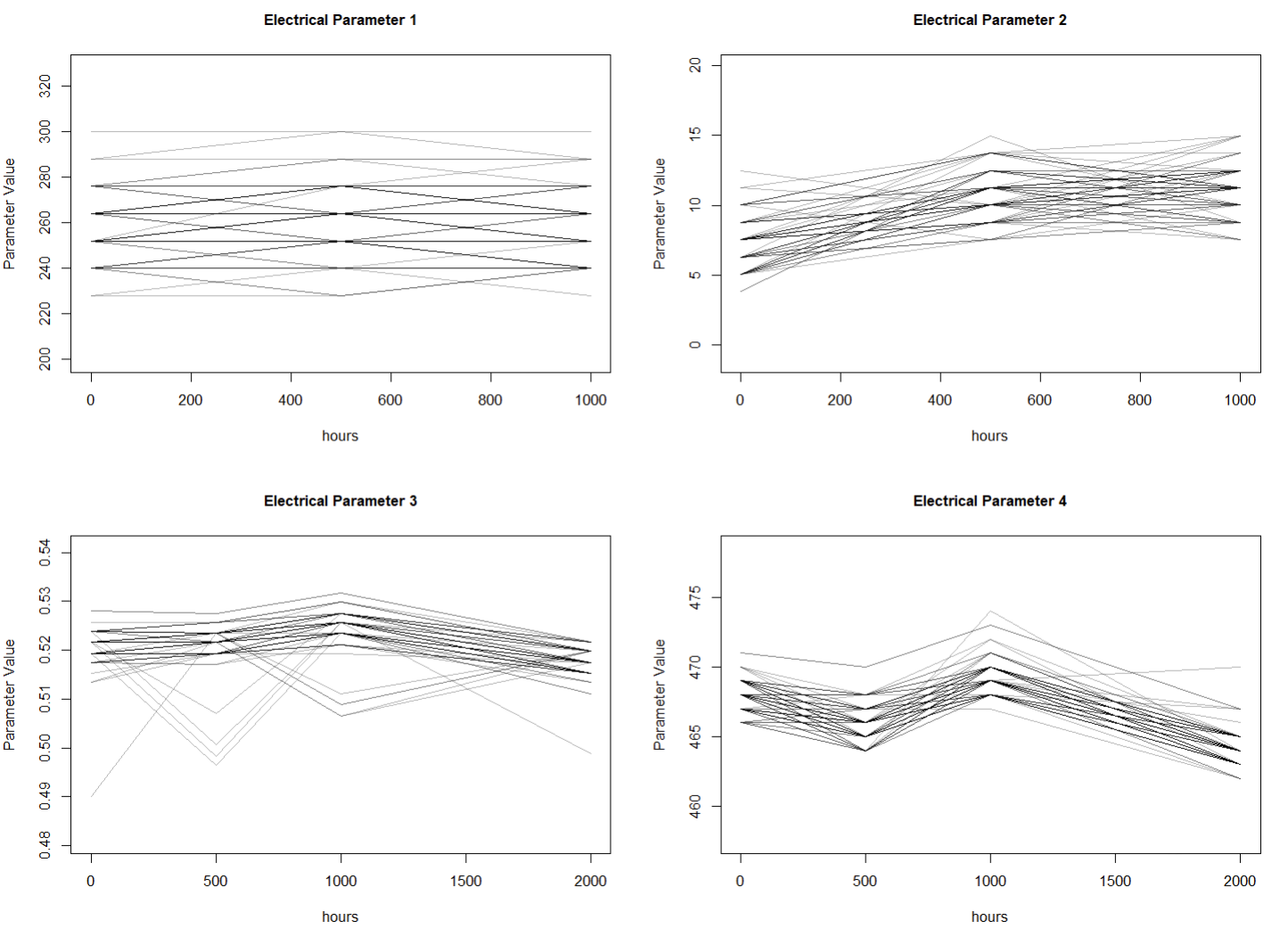}
	\caption{A set of anonymized real datasets used to illustrate possible behaviours of stress test panel data. Shown are constant drift, slight upward trajectories, grouping formation and a changing trajectory behavior.}
	\label{example_data}
\end{figure}

This study focuses on discrete parameters and modeling of  lifetime drift in accelerated stress tests, excluding temperature drift. The data analyzed is a sample of anonymized real data from rapid product testing (RPT) stress tests at Infineon Technologies Austria AG, as shown in Figure \ref{example_data}. The accelerated stress test data set contains $80$ devices for each electrical parameter, measured at $3$ different time-points. The anonymized data was transformed to have a step-size of 1 and the time-steps were normalized to $100$ hours each. The data exhibits several issues, including non-homogenous variance behavior, non-symmetric transition probabilities, and discrete data coinciding with integer values. To address these issues, the data standardization is necessary.

For step-size, first, we isolate the actual discretization of the data, represented by the least common multiple of the differences between data points. Then, we determine the offset of the data to the integers. Formally, $x_{i, centered}= \frac{x_i-min(x_j)}{lcm(\Delta x_{k})},\ j=1,...,n;\ k =1,...,n-1$, where $\Delta x$ denotes the differences in the real data points, $n$ the number of trajectories, and $lcm$ the least common multiple function. This ensures that the data $x_{centered}$ coincide with the integers $\mathbb{Z}$, allowing for standard notation to be applied. The translation by the minimum of all data points guarantees that data points are close to 0, preventing the state space from consisting of unnecessarily large integers. All results can be transformed back to the original discrete state space after calculation using the backtransformation $x_{i, centered}\cdot lcm(\Delta x_{k}) + min(x_j)= x_i,\ j=1,...,n,\ k =1,...,n-1$. From this point on, all formulae assume the state space to remain in the integers.

\subsection{The Structure of Stress Test Data}
Discrete accelerated stress test data are time-wise interval-censored time series with a discrete state space. The readout times are denoted by $t_0, \dots, t_k$, and the readout values are represented by $x_{i,0}, \dots, x_{i,k}$, indicating the parameter values of the $i^{th}$ device, with $i=1, \dots, d$, at times $t_0, \dots, t_k$. For instance, there may be 77 parts tested at 0, 168, 500, and 1000 hours. In general, the time differences may be arbitrary. Models for lifetime reliability must be flexible enough to deal with a wide variety of time differences.

\subsection{Correction for Tester Offset}
In accelerated stress tests, parameters are measured before and during the test, which may involve different sets of equipment and operators. This leads to a bias known as the tester offset, which must be addressed separately. To correct the data for the tester offset, unstressed reference devices (comparison parts) are measured alongside the stressed devices at each readout time. The drift observed in the unstressed parts is subtracted from the parameter values of the stressed parts to obtain the true drift.

\begin{figure}
	\centering
	\includegraphics[width=.9\linewidth]{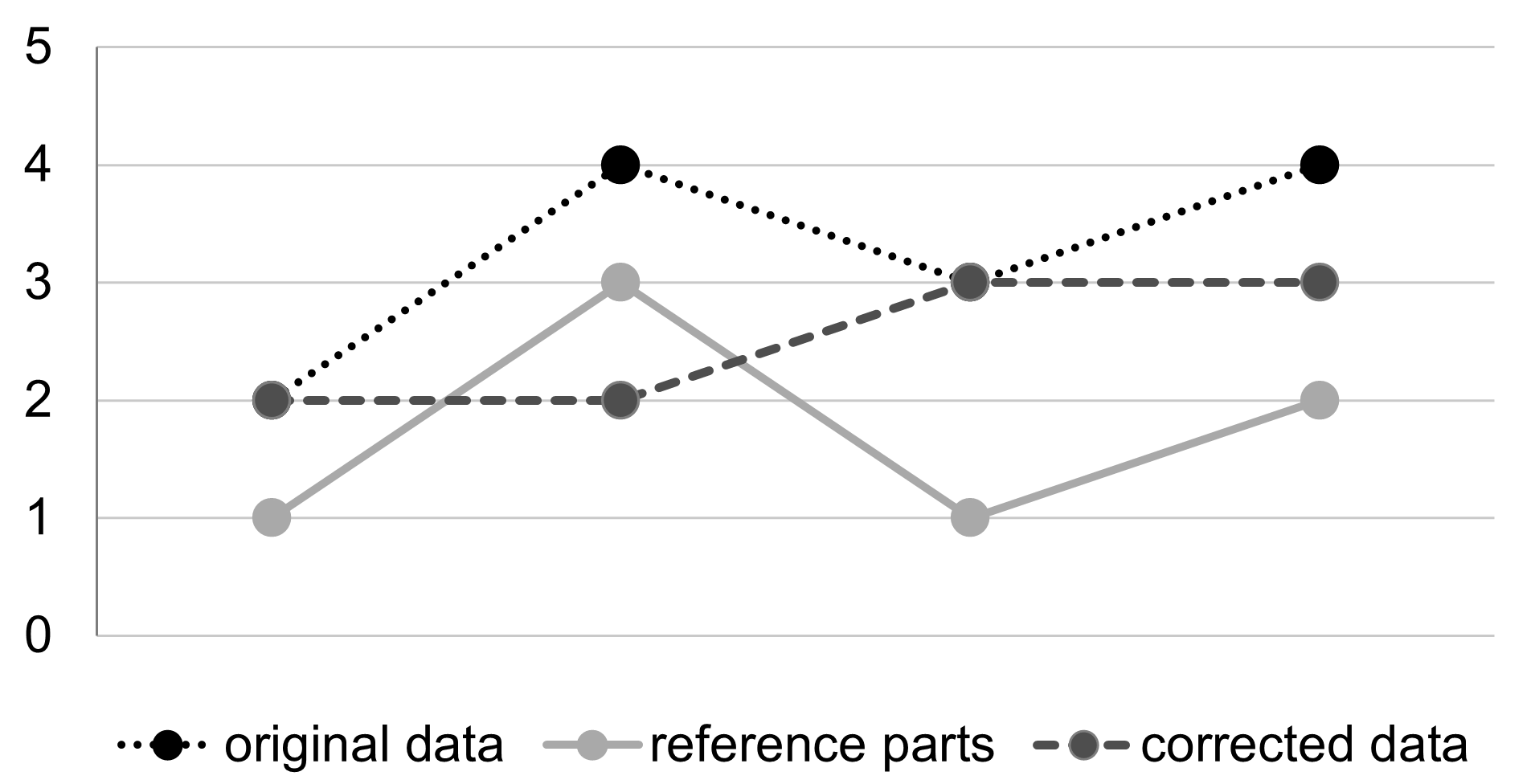}
	\caption{Removing the tester offset leads to smaller drift. It can be seen that the upward drift in original data diminishes greatly when corrected by reference parts.}
	\label{fig:testoffset}
\end{figure}

Let $c_{i,k}$ be the values of the $i=1, \dots, v$ comparison parts measured at readout $t_k$ and $x_{j,k}$ $j=1, \dots, d$ the measured values of the parameter of the stressed devices at readout time $t_k$. The offset-corrected stressed device values are calculated as:
\begin{align}
	x_{i,k,\text{offset-corrected}}=x_{i,k}-(\bar{c_k}-\bar{c_0}),
\end{align}
where $\bar{c}$ represents the arithmetic mean of the parameter values of the comparison devices. Figure \ref{fig:testoffset} shows an example of tester-offset-corrected data. For a more detailed derivation of the formula, refer to \cite{SommereggerPaper} and \cite{Hofer}. In practice, using the median instead of the arithmetic mean of the comparison part drift is often recommended to obtain a more robust correction, as outliers in reference devices can significantly skew the mean, especially when there are only a few reference devices.

\section{A Markov Model for Discrete Life-time Drift}\label{model}
We propose a transition model for total lifetime drift, where the total drift is modeled based on one-step transition matrices containing information about conditional changes. This approach addresses issues of skewed conditional distributions through non-parametric density estimation of conditional probabilities and maintains a simple optimization formulation. In the multivariate case, a Markov chain of order $n$ is constructed, and a Bayesian update is performed on the error term to update the conditional distribution dependent on the $n$ latest time-steps.

As the data set for typical stress test data is too small to directly and meaningfully estimate transition probability distributions ($n=77$), we use pooled increment data to estimate the error term distribution of a linear model for correlation effects to arrive at a non-parametric probability mass estimation for the conditional distribution at each readout depending on the previous state.

\subsection{Model Formulation}
The lifetime drift of the parameters of the empirical sample of stress test is assumed to represent the drift behavior of future products. However, the initial distribution may vary between wafers (hundreds of individual chips) and lots (production batches containing multiple wafers). Therefore, it is important to separate the model for lifetime drift from the model for the initial distribution of parameters. We consider the random variable containing total parameter drift up to time $t_k$ to be the sum of individual drifts between readout times.
\begin{align}
	D_{\Sigma_k}&=D_1+D_2+...+D_k. \nonumber\\
	&=(X_1-X_0)+(X_2-X_1)+\dots+(X_k-X_{k-1}).
\end{align}
The initial distribution can then, if desired, be considered separately:
\begin{align}
	X_k&=X_0+D_{\Sigma_k}
\end{align}

\subsection{A Markov Model}
The random total drift $D_{\Sigma_i}, \ i=1,...,k$ is modeled as a Markov chain process \cite{mchain}. This implies that dependencies are considered only between adjacent readout times, such as $t_1$ and $t_2$, but not $t_1$ and $t_3$. For higher dimensional dependencies, Markov chains of arbitrary order can be considered. To maintain the state space for all matrices, the total drift is directly modeled as the sum of individual drifts. Each Markov chain is uniquely determined by its transition matrices $T_i, \ i=1,\dots,k$.\\

$T_i$ is a stochastic matrix \cite{MarkovChains}, where each entry in $T_{ji}$ corresponds to the transition probability between states $i$ and $j$. The state space $S$ has a cardinality $\alpha$, and each row of $T$ sums to a total probability of one. This describes a conditional probability mass function, conditioned on the initial state denoted by the row.

\begin{align}
	P(D_{\Sigma_k}=j|D_{\Sigma_{k-1}}=i)=T_{ji}, \quad \forall (j,i)\in S\times S\\
	\sum_{j= 1}^{\alpha}T_{ji}=1,\quad \forall i =1,\dots,\alpha.
\end{align}
This stochastic matrix is then also called a Markov transition matrix.
The probability mass function of the parameter at the $i^{th}$ readout can then be written as 
\begin{align}
	p_{D_{\Sigma_k}}=e_0 \prod_{i=1}^k T_i.
\end{align} 
$e_0$ describes the canonical vector with a $1$ at the position corresponding to the state $0$ in $S$, $e_0=(0,\dots,1,\dots,0)$. This denotes a starting drift of $0$, i.e., a centering of total drift and independence of initial distribution.\\

\subsection{Estimation of Transition Probabilities}



To describe the transition matrices, we need transition probability densities $p_{D_{\Sigma_2}|D_{\Sigma_1}}$ for each possible starting location at $t_1$. However, in practice, with limited data ($n<100$), direct estimation can be challenging. To address this issue, we assume homogeneity of conditional distributions, allowing us to pool the data over all initial values at the same time-step and obtain a meaningful estimation of the conditional distribution shape.\\
The pooled increment distribution $p_{I_k}$ is considered representative of the conditional increment distributions, with some linear transformation $l$ affecting only location and scale.\\
An additional assumption is a linear dependence effect between readouts, which allows for modeling both positive and negative dependence. This helps prevent the propagation of uncertainty via the introduction of correction of variance via negative dependence. We model the dependence behavior as the random total drift $D_{\Sigma_k}$ having a fixed expected value and a linear dependence on $D_{\Sigma_{k-1}}$ plus some error term $\epsilon_k$, which may not necessarily be Gaussian.\\

The shape of the pooled increment distribution is taken as representative of the shape of the conditional distribution. For location and scale parameters, we consider the following model to describe the dependence:
\begin{align}\label{linmod}
	D_{\Sigma_k} = \beta_{0,k} + \beta_{1,k} \cdot D_{\Sigma_{k-1}} +\epsilon_k.
\end{align}
From this we obtain the conditional mean and variance parameters for $D_{\Sigma_k}|D_{\Sigma_{k-1}}$:
\begin{align}\label{mu}
	E(D_{\Sigma_{k}}|D_{\Sigma_{k-1}}=j)=\beta_{1,k} \cdot (j-E(D_{\Sigma_{k-1}})) + E(D_{\Sigma_{k}}),
\end{align}
\begin{align}\label{sigma}
	Var(D_{\Sigma_{k}}|D_{\Sigma_{k-1}}=j)=Var(D_{\Sigma_{k}}|D_{\Sigma_{k-1}})=Var(\varepsilon_k)=\sigma^2_{\varepsilon_k}.
\end{align}



The estimation of transition probabilities involves determining the conditional distribution shape of the error term, which is carried out through non-parametric means. This process involves estimating the pooled residuals along the linear model of dependence. Alternatively, depending on expert knowledge or stronger assumptions, the pooled increments $D_k:=D_{\Sigma_{k}}-D_{\Sigma_{k-1}}$ can be used as a representative for the final conditional distribution shape.\\

It is important to note that when pooling residuals, the resulting pooled data may not be in the correct discrete state space. For instance, in a linear model applied to integers, the residuals are generally not integers. In the case of pooled increments, the increment distribution is typically wider than the distribution of an error term in a conditional model. Therefore, when pooling increments, the variance and mean of the distribution need to be corrected to the conditional mean and variance as expressed in Equation \ref{mu} (denoting $\mu$) and Equation \ref{sigma} (denoting $\sigma$). In the case of residual estimation, the distribution must be made to conform to the integer state space. This is achieved by combining a continuous kernel density estimation \cite{KDE} of shifted and scaled data with piece-wise integration, or a re-binning function. The kernel density estimation estimates the arbitrary error term distribution with a continuous distribution, which is then discretized with the re-binning function. The definition of kernel density estimation is given in the Appendix in Definition \ref{def:KDE}.\\

The choice of kernel shape is a hyper-parameter and can in principle be chosen arbitrarily. In the case that the kernel density function is chosen as being rectangular with a width of $1$, the above expression reduces to the empirical histogram estimation, or the density function of the discrete data.
The resulting continuous distribution is then transformed to fit on the discrete state space $\mathbb{Z}$ with a binning function. The definition of a binning function is given in the Appendix in Definition \ref{def:Bin}	

In the case of increment pooling, the parameters for the linear transformation arise from Equations \ref{linmod}-\ref{sigma} and serve to transform the probability mass function $p_{I_k}$ to the location and scale parameters of $p_{D_{\Sigma_{k}}}$.	We denote these correction parameters with $\alpha_{k,1}$ and $\alpha_{k,2}$, respectively:
\begin{align}
	\alpha_{k,1}(j):=E(D_k)-(\beta_{1,k} \cdot (j-E(D_{\Sigma_{k-1}})) + E(D_{\Sigma_{k-1}}))
\end{align} is used for the location and
\begin{align}
	\label{alpha2} \alpha_{k,2}:=\sqrt{\frac{\sigma^2_{\epsilon}}{\sigma^2_{D_{k}}}}
\end{align}
is used for scale, where $\sigma^2_{\epsilon}$ is the variance of the residuals of the linear model and $\sigma^2_{D_k}$ is the variance of the pooled increments. Only the location correction parameter is dependent on the state $j$ as only the conditional expectation depends on the previous state , as in Equation \ref{mu}.

In total, the conditional probability mass function can be written as

\begin{align}
	\widehat{p}_{D_{\Sigma_{k}}|D_{\Sigma_{k-1}}}(D_{\Sigma_k}=x)=\int_{x-0.5}^{x+0.5} \frac{1}{nh} \sum_{i=1}^n K\Big(\frac{\zeta-d_{k}^{i}}{h}\Big)  \ d\zeta, \quad x\in S.
\end{align}
with $d_k^{i}$ being the $k-th$ pooled empirical increments, linearly transformed to fit to the residual conditional distribution.
\begin{align}
	d_{k}^{i}=\frac{(x^{i}_{k}-x^i_{k-1})-\alpha_{k,1}(j)}{\alpha_{k,2}}.
\end{align}
Alternatively, we can write the kernel density function as $k_h(x)$ and the binning function as $b(x)$, and obtain
\begin{align}
	\widehat{p}_{D_{\Sigma_{k}}|D_{\Sigma_{k-1}}}=b \left( k_h \left( d_{k}^{i}\right)\right),
\end{align}
Note that in the case of residual pooling instead of increment pooling, only re-transformation to the integer state space has to be performed: $\widehat{p}_{D_{\Sigma_{k}}|D_{\Sigma_{k-1}}}=b \left( k_h \left( \epsilon_{k}^{i}\right)\right)$.\\

\begin{figure}
	\centering
	\includegraphics[width=\linewidth]{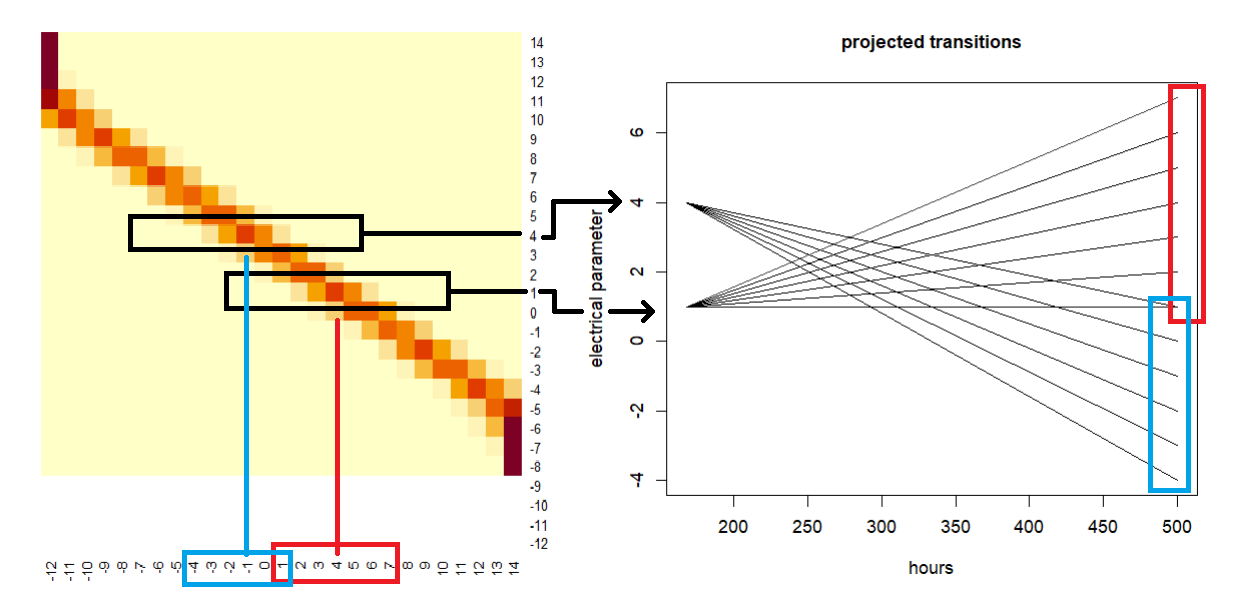}
	\caption{An illustration of a heat-map of a transition matrix and the matrices effect on the dependence of subsequent readout states. The transition matrix denotes the probabilities of change from the state denoted in the row to the state denoted by the column in each entry. Darker colors mean higher probabilities. In this case, the conditional distribution is linearly dependent on the starting position with homogeneous variance.}\label{Expl}
\end{figure}

\begin{figure}
	\centering
	\includegraphics[width=.8\linewidth]{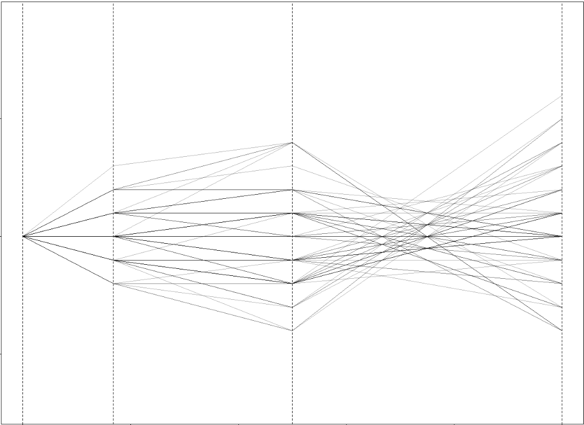}
	\caption{A set of synthetic electrical parameter lifetime trajectories centered at 0. The data show independent change between the first two readouts, positively correlated change between readouts 2 and 3 and negatively correlated change between readouts 3 and 4. Transition Matrices corresponding to drift behavior can be found in Figure \ref{Exp3}.}\label{Exp2}
\end{figure}

\begin{figure}
	\centering
	\includegraphics[width=.9\linewidth]{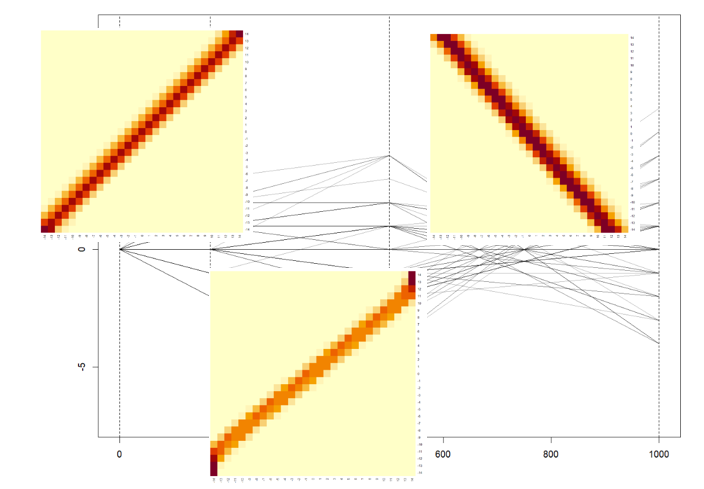}
	\caption{The estimated transition matrices corresponding to the first, second, and third time-step in the data set in Figure \ref{Exp2}. Note the diagonal matrix in case of independence and different tilts in case of positive or negative correlation. The model is linear in the conditional expectation and constant in conditional variance.}\label{Exp3}
\end{figure}

The transition matrix $T_k$ then has the entries 
\begin{align}
	T_k^{ij}=\widehat{p}_{D_{\Sigma_{k}}|D_{\Sigma_{k-1}}}(D_{\Sigma_{k}}=j,D_{\Sigma_{k-1}}=i),
\end{align}
which fall along a line determined by the correlation effect between the two readout points and variation along the line with a shape given by the pooled increments or residuals. An example of such a transition matrix containing the conditional distributions can be seen in Figure \ref{Expl}. The effect of different positive and negative dependence in the discrete case on the shape of the transition matrix can be seen in Figures \ref{Exp2} and \ref{Exp3}.

In the case of the first drift, $D_{\Sigma_{1}}$, we assume independence of $X_0$, and no correlation correction is needed in either case, as we assume homogeneity in conditional variance:
\begin{align}
	T_1^{ij}=\widehat{p}_{D_{\Sigma_{1}}|X_0}(D_{\Sigma_{1}}=j,X_0=i)=\widehat{p}_{D_{\Sigma_{1}}}(D_{\Sigma_{1}}=j).
\end{align}

The resulting process estimation can be viewed as similar to fitting a discrete AR(1)-model with non-Gaussian residuals between each two available readout times. The full model is then a chain of these different AR(1)-like-steps with different parameters. For in-between time-steps estimation, the same process of estimating densities can be used on the interpolated empirical or model values. This allows arbitrary stopping times and combination with usage profiles that denote different weights of usage to different lifetime intervals in vehicles.

\subsection{Model Limitations and Assumptions}
\label{sect:limits}
The model has several limitations. As shown in Section \ref{sect:complexity}, the optimization of guard bands is a problem of complexity $O(n^5)$ in the worst case. Although it may be faster than comparable methods in small state spaces, in the case of huge state spaces, a more efficient implementation may be necessary for it to compete with faster implementations of state-of-the-art methods. Furthermore, the transition model only considers single step dependence structures. This could be extended in future work to multi-step dependence structures. Furthermore, the model is not accounting for strong group effects in data. If the discrete data additionally shows strong grouping effects, the model may become inaccurate with respect to real data behavior. This, too is a topic for future research.\\

Regarding assumptions, the proposed transition model is built upon two fundamental assumptions: the linearity of dependence of the conditional expectation between adjacent readout points and the homogeneity of variance of the conditional distribution. While these assumptions are crucial to the model's validity, they are inherently difficult to verify in real data. The linearity of expectation can be regarded as a first-order Taylor approximation of the true underlying dependence function, thereby representing a simplifying assumption. Likewise, the homogeneity of the conditional variance can be seen as a simplification of the true underlying conditional variance function, which, in principle, could be modeled with greater complexity. However, the sparsity of data in real stress test sample data sets renders it almost impossible to meaningfully estimate the conditional variance for every initial state, thereby necessitating these assumptions.

\section{Guard Banding}\label{gbs}
\begin{figure}
	\includegraphics[width=\linewidth]{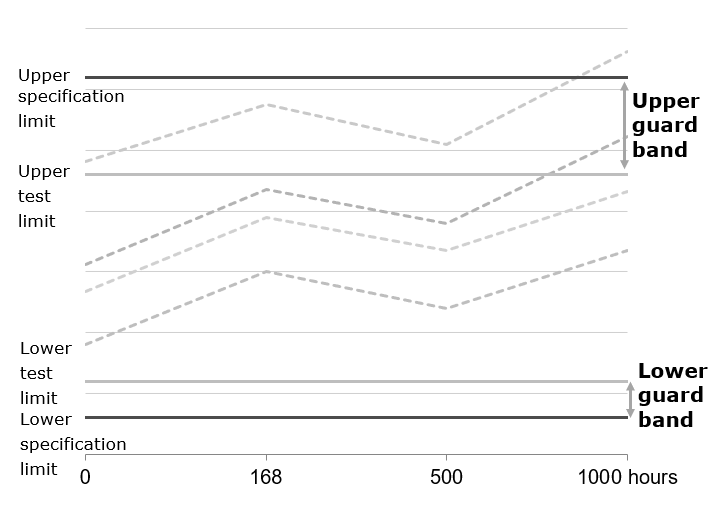}
	\caption{An example of guard banding - the guard band prevents devices from starting too high so that they eventually exceed the specified limits.}\label{GB}
\end{figure}

Guard bands aid in ensuring that the electrical parameters of devices do not overstep their specified limits within usage during their lifetime. A guard band is the area between pre-specified limits which guarantee functionality and tighter test limits at the last testing before shipping, which are introduced for quality control purposes. Parts with parameters outside those test limits are deemed too risky and will not be shipped, thus guaranteeing quality to the customer.
This is shown in Figure \ref{GB}. 

\subsection{Definition of Guard Bands for Electrical Parameters} 

Both guard bands and limits may be defined as one-sided, that is, only upper or lower limits are specified, which may be useful in the case of physical processes where exceedance only makes physical sense in one direction, or two-sided, which means that both upper and lower limits are relevant. Guard bands in general are used to control for temperature drift, measurement uncertainty, and lifetime drift. We focus on the part concerning lifetime drift separately.

The optimization problem of guard banding lies in minimizing the lifetime exceedance probability of specification limits up to a certain quality target while maximizing yield, that is, minimizing the amount of parts falling inside of guard bands.
A guard band is the area between a specified test limit and a tighter limit at production testing. E.g., for an upper specified limit $USL$ and an upper test limit $UTL$ the upper guard band $USL-UTL$ is written as $GBU$. The complete definition of a guard band is given in definition \ref{def:GB} in the Appendix.
	The final production test is a 100 \% control measure that ensures that all shipped devices are within the tighter test limits.

	Given a quality target $q$ of $1$ppm, or one in a million devices allowed to overstep the limits in their lifetime, the likelihood to stay within the specified limits over lifetime can be written as, see also \cite{Hofer},
	
	\begin{equation}
		P\left(X_1,\ldots,X_m \in \left[LSL,USL\right] \mid X_0 \in \left[LTL,UTL\right] \right) \geq q.
		\label{PoutsideLimits}
	\end{equation}


\subsubsection{For One-Sided Guard Bands}

For one-sided guard bands, exceedances in only one direction, above or below, are considered exclusively. The respective other limit is considered fixed. This may be helpful in cases where drift can occur in only one direction. The formula is then given in Equation \ref{condit_up}, also discussed in \cite{Hofer} and \cite{SommereggerPaper}.

The optimization problem for upper test limits is
\begin{align}
	&max. \ UTL\ s.t. \nonumber\\ 
	&\sum_{k=1}^{m} P(X_k>USL \bigcap_{0<j<k} X_j \leq USL\mid X_0 \leq UTL)\leq 1-q, \label{condit_up}
\end{align}
with $UTL\leq USL$ and $\bigcap_{0<j<1}$ denoting the nullary intersection.

The equation for lower test limits can be written analogously.
\subsubsection{For Two-Sided Guard Bands}

For two-sided guard bands, both upper and lower limit exceedance probabilities are considered at the same time. This means that the optimization problem becomes two-dimensional and therefore, harder to solve. In particular, the uniqueness of a solution is no longer guaranteed. We therefore introduce another constraint to the maximization problem, in this case, maximization of yield, or shipped product to guarantee uniqueness in the solution. This corresponds to a maximization of $UTL-LTL$. 

The equation to solve is, see also \cite{Hofer}:
\begin{align}
	&\max_{LTL,UTL}(UTL-LTL) \ s.t. \nonumber \\
	&\sum_{k=1}^{m} P\left(\left((USL < X_k) \cup (X_k < LSL)\right) \bigcap_{0<j<k}LSL \leq X_j \leq USL\mid LTL \leq X_0 \leq UTL\right) \nonumber \\
	& \leq 1-q, \label{condit_two}
\end{align}
with $LSL\leq LTL<UTL\leq USL$.

\subsection{Calculation of Guard Bands for Single Devices using Discrete Markov Chain Modeling} \label{singledevsect}
Guard bands and quality targets can be calculated in two ways. The first way is to specify a quality target based on single devices. That means that every produced device may have at most a chance of $1-q$ of drifting outside of the specified limits over lifetime. The other concerns are the quality target applied to a specific batch of devices with known initial distribution. In this section, we will show how to use the discrete transition model to formulate the guard band problem combined with a given distribution for the initial state vector.

\subsubsection{For One-Sided Guard Bands}

For one-sided upper guard bands, in a transition matrix notation, the optimization of guard bands with regard to yield seeks to find the best upper test limit ($UTL\in S$) that guarantees lifetime quality. \begin{align} \label{within}
	&	max(UTL)\nonumber\\
	&	s.t. \nonumber\\
&	\sum_{k} P(X_k\leq USL| X_{0<i<k}\leq USL,X_0\leq UTL)\cdot P(X_{0<i<k}\leq USL\cap X_0\leq UTL) \geq q.
\end{align} This corresponds to the probability of remaining inside the limits satisfying the reliability criterion $q$. 
The above conditional expression 
can be represented by the transition matrix $T_{k-i}$ where all values that denote $P(X_{k-i} > USL)$ and $P(X_{k-i-1} > USL)$ are set to 0, which means that the entries of all rows and columns denoting states $s\in S$ for which hold $s>USL$ are set to $0$. These restricted transition matrices will be abbreviated with $T_i|_{(a,b)\times(c,d)}$ denoting that all entries outside of the rows denoting states $a$ to $b$ and outside of all columns denoting states $c$ to $d$ are set to $0$. The vector $\mathbf{v_k}$ contains the summed probabilities of conditional distributions (Eq.\ref{within}). $e_0$ describes the canonical vector with a $1$ at the position corresponding to the state $0$ in $S$, $e_0=(0,\dots,1,\dots,0)$. The optimization problem can be formulated as follows:
\begin{align}
	&max(UTL) \ s.t. \nonumber \\
	&q \leq \sum_{i} v_k^i \nonumber\\
	&\mathbf{v_k}=\sum_{k} \mathbf{e_0}\cdot \prod_{t=1}^k T_t|_{(USL-UTL,LSL)\times(USL-UTL,LSL)}, \label{optsingleu}
\end{align}

For lower test limits, the equation can be written analogously to Equation \ref{optsingleu}.

\subsubsection{For Two-Sided Guard Bands}
For two-sided guard bands, parameters may drift outside both upper and lower specified limits. In that case, to guarantee the uniqueness of the solution, the yield maximization criterion is introduced. $e_0$ describes the canonical vector with a $1$ at the position corresponding to the state $0$ in $S$, $e_0=(0,\dots,1,\dots,0)$.
\begin{align}
	&max(UTL-LTL) \ s.t. \nonumber \\
	&q \leq \sum_{i} v_k^i \nonumber\\
	&\mathbf{v_k}=\sum_{k} \mathbf{e_0}\cdot \prod_{t=1}^k T_t|_{(USL-UTL,LTL-LSL)\times(USL-UTL,LTL-LSL)}, \label{optboth}
\end{align}

\subsection{Calculation of Guard Bands with Initial Distribution} \label{initdevsec}
If we wish to apply the quality target for a specific batch with an arbitrary initial probability mass function, the lifetime drift is combined with the probability mass function of the parameters at production testing. Let $\mathbf{d}|_{(UTL, LTL)} $ be the vector containing the initial probability mass function of the parameter, with values above $UTL$ and below $LTL$ set to $0$. Then, we can calculate, both one-sided and two-sided guard bands.
\subsubsection{For One-Sided Guard Bands}
For the one-sided upper guard band including the initial probability mass function, the optimization problem to solve is
\begin{align}
	&max(UTL) \ s.t. \nonumber \\
	&q \leq \sum_{i} v_k^i \nonumber\\
	&\mathbf{v_k}=\sum_{k} \mathbf{d}|_{(UTL,LSL)} \cdot \prod_{t=1}^k T_t|_{(USL,LSL)\times(USL,LSL)}. \label{optinitup}
\end{align}
For the one-sided case for lower guard bands considering the initial probability mass function, the equation can be written analogously to Equation \ref{optinitup}.

\subsubsection{For Two-Sided Guard Bands}
For the calculation of two-sided guard bands, the calculation includes cases of exceedances above and below, 
\begin{align}
	&max(UTL-LTL) \ s.t. \nonumber \\
	&q \leq \sum_{i} v_k^i \nonumber\\
	&\mathbf{v_k}=\sum_{k} \mathbf{d}|_{(UTL,LTL)} \cdot \prod_{t=1}^k T_t|_{(USL,LSL)\times(USL,LSL)}. \label{optinit}
\end{align}
An example workflow containing pre-processing, modeling and optimization can be found in the Appendix in Figure \ref{fig:wf}.

\subsection{Practical Relevance}
The automated modeling of lifetime drift in electrical parameters using accelerated lifetime stress test data offers several significant advantages in industrial semiconductor manufacturing. By enabling automatic control of lifetime quality behavior, this approach can ensure the reliability of device types. Moreover, the specification of parameters by engineering or quality departments in stress tests often results in the measurement of thousands of electrical parameters per device. A statistical model can be leveraged by quality engineers to set test limits that optimize yield while guaranteeing quality.\\
Furthermore, a statistical model for lifetime drift facilitates the identification of critical parameters that indicate degradation in semiconductor chips. These parameters can be monitored in real-world applications, such as autonomous vehicles, to estimate the remaining useful life of the chip. This enables predictive maintenance actions to be taken by decision systems within the vehicle, thereby preventing potential breakdowns.\\
Finally, the detection of higher-than-usual guard bands in stress test data may indicate deviations in productive processes that are not apparent in process control. This can lead to the early detection of manufacturing deviations, resulting in improved yield and product quality.

\section{Comparison to State-of-the-art and Simulation}
\label{Comparison}
Comparing our approach to other models proves challenging due to the distinct objectives and assumptions underlying each methodology. While panel data analysis typically aims to isolate treatment effects in different groups of data, our focus lies in modeling the probability distribution of the entire sample.\\
Existing continuous multivariate approaches to longitudinal data, such as those presented in \cite{Liang,lgpr} and \cite{TLongMiss}may require adaptations to accommodate discrete data. However, even with such adaptations, discretization errors can propagate and significantly impact tail distribution behavior, depending on the data resolution.\\
More flexible models, including copulas, see \cite{Sklarcop}, and Markov-chain Monte Carlo (MCMC) approaches, are computationally intensive and therefore unsuitable for large-scale industrial applications and edge devices. Non-parametric and transition-based approaches for multivariate longitudinal data typically focus on continuous data, as seen in \cite{MLNP}. Moreover, Sklar's theorem does not hold for discrete data, rendering classical copula approaches inapplicable, see \cite{Geenens}. The only readily available comparison method to the authors is the method developed in \cite{SommereggerPaper} for continuous data.\\

\begin{figure}\centering
	\includegraphics[width=1.1\linewidth]{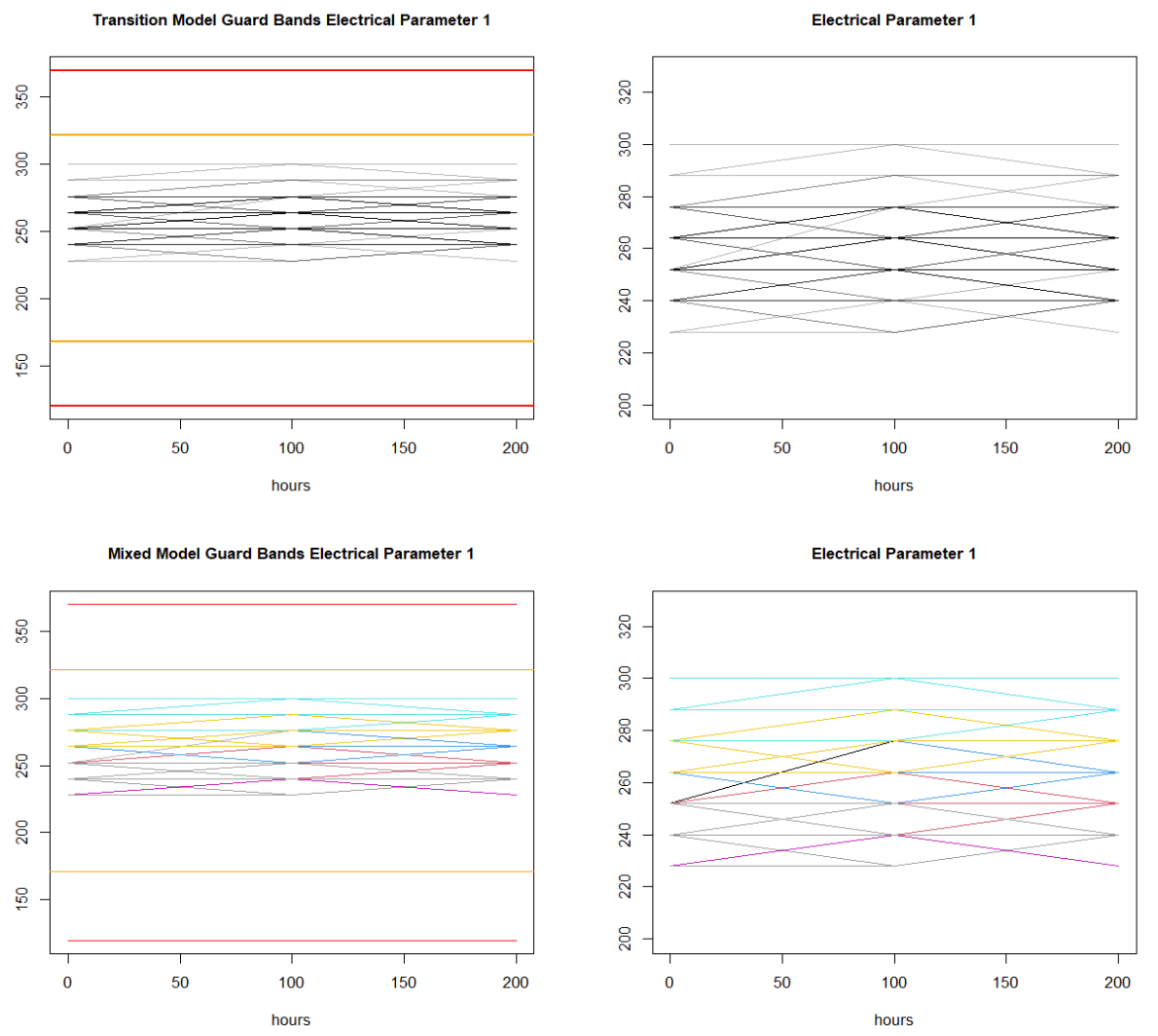}
	\caption{Results for the guard bands for the Electrical Parameter  trajectory data set. Models are the transition model proposed and the the multivariate Gaussian model from \cite{SommereggerPaper}. On the left is the visualization with specified limits, on the right only the data. Each model cluster is denoted with a different color. Specified limits are given in red, calculated test limits in orange.}
	\label{example_data2}
\end{figure}

In this section, we will demonstrate the calculated guard bands on sets of (anonymized) real data from semiconductor manufacturing. The anonymized data sets used can be seen plotted in Figure \ref{example_data}. A result from a state-of-the art algorithm from \cite{SommereggerPaper} can be seen in Figure \ref{example_data2}. The clustering behavior of discrete data patterns can be observed well.
Furthermore, we will assess the capability of the model to accurately replicate synthetic patterns of discrete data that may arise in real-world productive data, seen in Figure \ref{fig:synth}.

\subsection{Comparison to Multivariate Gaussian Model}

We compare guard bands calculated from the continuous model from \cite{SommereggerPaper} with the results from transition model presented in this work. The upper and lower specified limits (USL and LSL) for the electrical parameters shown in Figure \ref{example_data} are given as in Table \ref{table:usl_lsl}:

\begin{table}[ht]
	\centering
	\begin{tabular}{|c|c|c|} 
		\hline
		Trajectories & USL & LSL  \\
		\hline
		\textbf{Parameter 1} & 120 & 370 \\ \hline
		\textbf{Parameter 2} & -300 & 1000 \\ \hline
		\textbf{Parameter 3} & -1.9 & 1.9\\ \hline
		\textbf{Parameter 4} & -1.9  & 1.9 \\ \hline
		
	\end{tabular}
	\caption[Values Table]{Table of specified limits for real data electrical parameters. USL/LSL denote upper/lower specification limit.}
	\label{table:usl_lsl}
\end{table}


In Table \ref{table:comparison_old_ours}, we compare results from using the mixed Gaussian modeling approach shown in \cite{SommereggerPaper} with the approach outlined in this paper. In the case of guard banding, smaller guard bands at the same quality level are an improvement. The quality level chosen was $1ppm$, the standard used in semiconductor manufacturing. 

The total time for calculation of transition matrices and guard band optimization on 13th Gen Intel(R) Core(TM) i5-1345U with 16 GB RAM took $22.70$ seconds for the transition model and $9.22$ seconds for the mixed Gaussian model.

\begin{table}[ht]
	\centering
	\begin{tabular}{|c|c|c|c|c|} 
		\hline
		Trajectories &  GBU-old & GBL-old & GBU-ours & GBL-ours \\
		\hline
		\textbf{Parameter 1}  & 48.881 & 51.216  & 48 & 48 \\ \hline
		\textbf{Parameter 2} & 15.763 & 7.782 & 16.133		 & 8.687
		\\ \hline
		\textbf{Parameter 3} &  0.0747 & 0.0562  & 0.040		 & 0.0276
		\\ \hline
		\textbf{Parameter 4} & 0.0104	 & 0.0148 & 0.009		 & 0.011
		\\ \hline
		
	\end{tabular}
	\caption[Values Table]{Table of calculated guard bands for real data electrical parameters. GBU/GBL denote upper/lower guard bands. Smaller bands are better with respect to yield.}
	\label{table:comparison_old_ours}
\end{table}

It can be seen that the discrete model leads to tighter guard bands results in three of four cases, namely electrical parameter 1 and electrical parameters 3 and 4. In the case of electrical parameter 2, although the data exhibits clear group formation and the discrete model does not take grouping into account directly, the results are still close to the result of the continuous model that explicitly takes into account grouping effects. By this, it can be seen that the non-parametric transition approach can even capture some grouping behavior without making assumptions on the distribution behavior of the groups. 

\subsection{Complexity of Modeling and Optimization}
\label{sect:complexity}
Given a state space of size $n$ and $k$ readout times, the generation of a transition matrix has the computational complexity of at least $O(n^2)$, as for each row, the transition probability of each state has to be calculated. This can be done in a number of operations independent of state space size, giving us an initial complexity of $O(k*n^2)$. For propagation, matrix multiplication has a complexity of $O(n^3)$ in the worst case. For the optimization, the naive grid search method uses another $O(n^2)$ operations. In total, the proposed method for optimizing guard bands has therefore an upper complexity bound of $O(k*n^2)+O(n^2)*k*O(n^3)=O(n^5)$. This stands in contrast to classical integer optimization problems which are NP-complete and therefore not solvable in polynomial time.

\subsection{Simulation}
\begin{figure}\centering
	\includegraphics[width=1.1\linewidth]{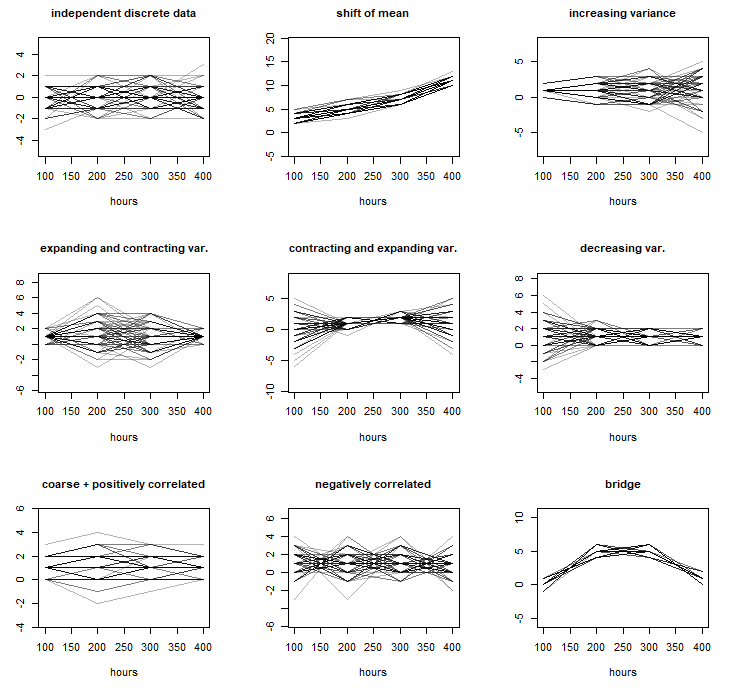}
	\caption{A set of synthetically created trajectory patterns. The data were generated in a way that a variety of patterns that may occur in real data sets are represented. The goal is to see if the data generated by the transition model can match a variety of behaviors.}
	\label{fig:synth}
\end{figure}

We assess the Markov chain model's capacity to accurately reproduce various patterns in discrete longitudinal data. To achieve this, we synthesized nine distinct patterns, each displaying a specific behavior in mean, variance, or correlation between time-steps, or a combination thereof. These patterns are depicted in Figure \ref{fig:synth} and consist of discrete trajectories with a quantization step size of $1$.

\begin{figure}\centering
	\includegraphics[width=1.1\linewidth]{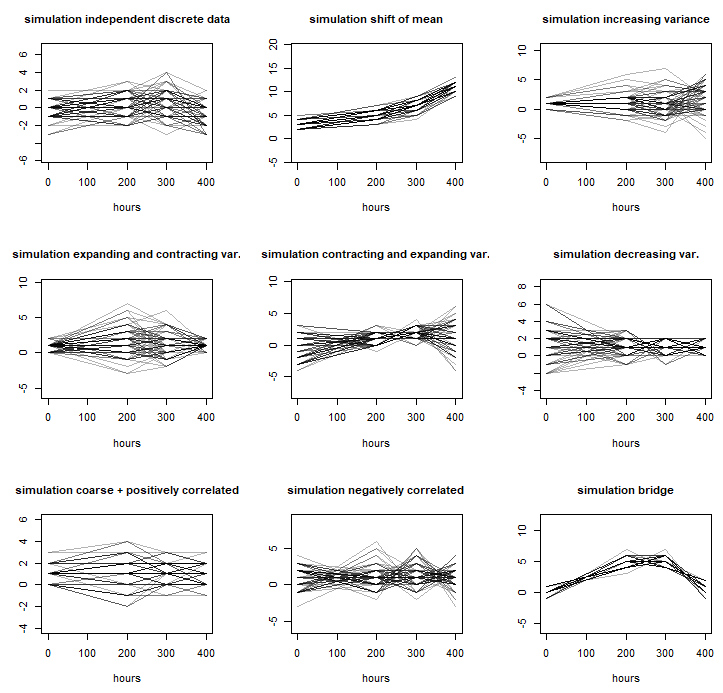}
	\caption{A set of simulated trajectory patterns based on transition matrices learned from the data in Figure \ref{fig:synth}, estimated via the stochastic Markov model. The generated data sets match the synthetically generated patterns well which shows promise that the model may be able to handle a variety of data behaviors often encountered in practice.}
	\label{fig:simul}
\end{figure}

We apply a Markov dependence model, as described in Section \ref{model}, to each of these patterns and utilize the resulting transition matrices to generate new simulations initiating from the original starting distribution. If the model successfully recreates the shape of the patterns, it may serve as an indicator of its potential usefulness when applied to real data that follow similar patterns. The outcomes of the simulation using the model are presented in Figure \ref{fig:simul}.

Overall, the model appears to effectively capture the diverse behaviors simulated. It successfully captures changes in variance over time, such as contracting and expanding variance, despite being based on conditional distribution estimation. Additionally, the model allows for the specification of dependence behavior in both positive and negative directions. Thanks to the non-parametric estimation method, the model is capable of capturing even slightly skewed distributions fairly well. In total, it can be said that the model is suitable for use in productive purposes in semiconductor manufacturing due to its statistical and therefore explainable nature and its ability to capture a wide variety of behaviors. Several limitations exist, which are covered in Section \ref{sect:limits} and suggested for further research in Section \ref{conclusion}.

\section{Conclusions}\label{conclusion}
In this work, we present a novel transition model for discrete longitudinal data, as generated in accelerated stress testing experiments within the semiconductor industry. We demonstrate how to use the model to fit arbitrary conditional distributions and employ the resulting transition matrix system to calculate optimal guard bands for quality purposes. The results are showcased on a set of real data from semiconductor stress testing and compared to an earlier method developed for purely continuous data. This work aims to fill the gap in industrially-usable methods for prognostics and health-management models concerning discrete data which remain data-driven, explainable, scalable and working on edge devices. The model is designed to capture arbitrary conditional distribution behaviors and can be extended to more complex formulations.

\section{Outlook} \label{sec8}
In future work, the model may be explored in various directions, including:
\begin{itemize}
	\item Hyper-parameter tuning, such as different kernel choices for the density estimation.
	\item Extension to discrete clustering and combination of Markov Models, along with the investigation of appropriate discrete clustering algorithms.
	\item Evaluation of the model's performance on different real datasets and comparison to other candidate lifetime drift guard banding models on a variety of data sets and state space sizes.
	\item Further mathematical investigation of higher-order Markov chains, asymptotic behaviors and extension of the model to use transition Tensors and variable conditional variance and expectation functions.
\end{itemize}

\section{Disclosure}
During the preparation of this work the authors used Llama2 in order to improve readability. After using this tool, the authors reviewed and edited the content as needed and take full responsibility for the content of the publication.

\section{Acknowledgments}
The project AIMS5.0 is supported by the Chips Joint Undertaking and its members, including the top-up funding by National Funding Authorities from involved countries under grant agreement no. 101112089.

\appendix




\newpage

\section*{Glossary}
\begin{longtable}{lll}
	$b(f)$&.....&binning function, applied to the function $f$.\\
	$c_k^j$&.....& value of the reference device $j$ at readout $t_k$.\\
	$\bar{c}_k$&.....&average of the reference devices at readout $t_k$.\\
	$D_k$&.....&difference of drift from readout $t_{k-1}$ to $t_{k}$, random variable.\\
	$D_{\Sigma_k}$&.....&total drift up to $t_{k}$.\\
	$e_{k}$&.....&$k^{th}$ canonical vector, vector containing only zeroes and a one at position $k$.\\
	$\hat{f}_h(x)$&.....&kernel density estimation function.\\
	$\omega_i$&.....&omega, mixture parameter in the multivariate case,\\&& steers dependence on past read-outs.\\
	$P(X)$&.....&probability distribution of the RV $X$.\\
	$p_X$&.....&probability density function of the RV $X$.\\
	$q$&.....&reliability target.\\
	$t_k$&.....&$k^{th}$ readout point at stress time $t$.\\
	$T_i$&.....&Markov transition matrix from time-step $i-1$ to time-step $i$.\\
	$T_i|_{(a,b)\times(c,d)}$&.....&Markov transition matrix restricted to the states $(a,b)\times(c,d)$. \\&&All entries outside of the given area denoted by row \\&&and column indices are set to $0$.\\
	$X_k$&.....&random variable of the value of an electrical parameter at readout $k$.\\
	$x^i_{k}$&.....&realization of the random variable $X_k$ at device $i$.\\
\end{longtable}

\section*{Abbreviations}

\begin{longtable}{lll}
	LSL&.....&lower specification limit.\\
	LTL&.....&lower test limit.\\
	ppm&.....&parts per million.\\
	s.t.&.....&subject to, denotes restrictions in optimization problems.\\
	USL&.....&upper specification limit.\\
	UTL&.....&upper test limit.\\
\end{longtable}

\section*{Data Processing and Modeling Workflow}
\begin{figure}\centering
	\includegraphics[width=1.1\linewidth]{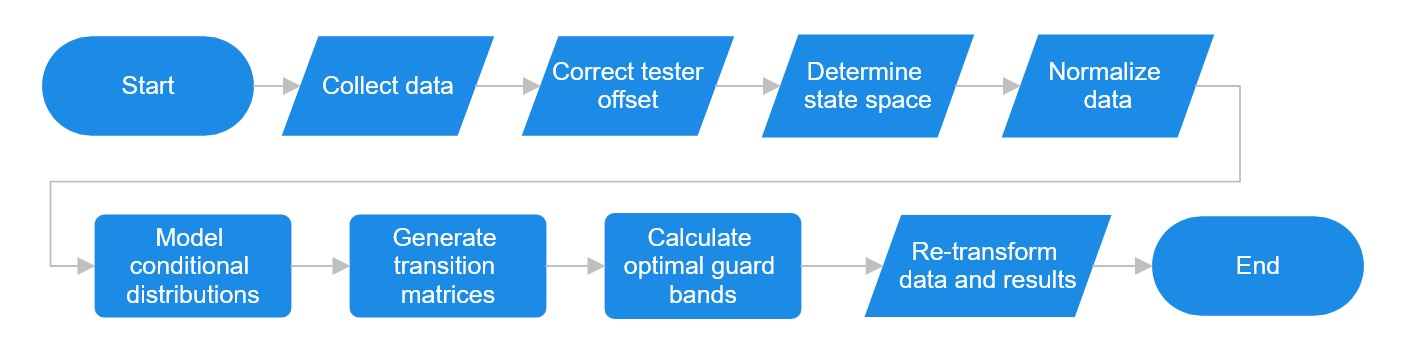}
	\caption{A workflow showcasing the steps necessary for obtaining optimal guard bands from discrete data via the transition model. First, preprocessing includes correcting and re-normalizing the data. Modeling can then be done in an easier way on normalized data. Then, the final results are re-transformed and used.}
	\label{fig:wf}
\end{figure} 

\section*{Definitions}
\begin{definition}[Kernel density estimation] 
	\label{def:KDE}
	Let $(x_1, \ldots,x_n)$ be i.i.d. samples from an underlying uni-variate distribution $f$, the kernel density estimator  \cite{KDE} of $f$ is then
	$$ \widehat{f}_h(x) = \frac{1}{n}\sum_{i=1}^n K_h (x - x_i) = \frac{1}{nh} \sum_{i=1}^n K\Big(\frac{x-x_i}{h}\Big),$$where the kernel $K$ is a non-negative function integrating to one and the bandwidth $h>0$ acts as a smoothing parameter. 
\end{definition}
\begin{definition}[Binning function]
	\label{def:Bin}
	A binning function is a function $b: f_{\mathbb{R}\rightarrow\mathbb{R}}\rightarrow g_{\mathbb{Z}\rightarrow \mathbb{R}}$ with $$(b\circ f)(x)=g(x),$$
	$$g(x)=\int_{x-0.5}^{x+0.5}f(\zeta) \ d\zeta, \quad x\in \mathbb{Z}.$$
	In particular, in the case of $f$ being a continuous density function, $b(f)$ gives the histogram of bin-width 1 centered on the integers.
\end{definition}
\begin{definition}[Guard Band]
	\label{def:GB}
	Let $X_t$ be a time series of an electrical parameter, $USL$ the upper specified limit, and $LSL$ the lower specified limit. Further, let $1-q$ be a quality target, e.g., 1 part per million affected, $1ppm=0.000001$ probability of drifting outside of limits.
	Guard bands are then the areas between tighter upper and lower test limits, $UTL$ and $LTL$, and the originally specified limits $USL$ and $LSL$, respectively.$$GBU=USL-UTL, \ GBL=LTL-LSL.$$ 
\end{definition}

 \bibliographystyle{elsarticle-harv} 
  \bibliography{references.bib}

\end{document}